\documentclass[aps,pra,twocolumn,floatfix]{revtex4-1}
\usepackage{graphicx}
\usepackage{dcolumn}
\usepackage{bm}
\usepackage{multirow}
\usepackage{ulem}
\usepackage{color}
\usepackage{longtable}
\usepackage{subfigure}
\usepackage{amssymb}
\usepackage{amsmath}
\usepackage{upgreek}
\usepackage{latexsym,epsfig}
\usepackage{float}
\usepackage[figuresright]{rotating}
\usepackage{upgreek}
\usepackage{ulem}
\usepackage{booktabs}

\begin{document}

	
\title{Application of General-order Relativistic Coupled-cluster Theory to Estimate Electric-field Response Clock Properties of Ca$^+$ and Yb$^+$}

\author{YanMei Yu$^{1,2}$} \email{Email: ymyu@aphy.iphy.ac.cn}
\author{B. K. Sahoo$^3$} \email{Email: bijaya@prl.res.in}
	
\affiliation{$^1$Beijing National Laboratory for Condensed Matter Physics, Institute of Physics, Chinese Academy of Sciences, Beijing 100190, China}
\affiliation{$^2$University of Chinese Academy of Sciences, 100049 Beijing, China}
\affiliation{$^3$Atomic, Molecular and Optical Physics Division, Physical Research Laboratory, Navrangpura, Ahmedabad 380009, India}
	
\begin{abstract}
Accurate calculations of electric dipole polarizabilities ($\alpha_d$), quadrupole moments ($\Theta$), and quadrupole polarizabilities ($\alpha_q$) for the clock states of the singly charged calcium (Ca$^+$) and ytterbium (Yb$^+$) ions are presented using the general-order relativistic 
coupled-cluster (RCC) theory. Precise knowledge of these quantities is immensely useful for estimating uncertainties caused by major systematic effects such as the linear and quadratic Stark shifts and black-body radiation shifts in the optical Ca$^+$ and Yb$^+$ clocks. A finite-field 
approach is adopted for estimating these quantities, in which the first-order and second-order energy level shifts are analyzed by varying strengths of externally applied electric field and field-gradient. To achieve high-accuracy results in the heavier Yb$^+$ ion, we first calculate 
these properties in a relatively lighter clock candidate, Ca$^+$, which involves similar clock states. From these analyses, we learned that electron 
correlation effects arising from triple excitations in the RCC theory contribute significantly to the above properties, and are decisive factors in bringing the calculated values closer to the experimental results.
\end{abstract}
	
\date{\today}
	
\maketitle
	
\section{Introduction}

The singly charged ytterbium ion (Yb$^+$) is a unique candidate, as three of its transitions, namely the $4f^{14}6s~^2S_{1/2} \rightarrow 
4f^{14}5d~^2D_{3/2}$ and $4f^{14}6s~^2S_{1/2} \rightarrow 4f^{14}5d~^2D_{5/2}$ electric quadrupole (E2) transitions and the $4f^{14}6s~^2S_{1/2} \rightarrow 4f^{13}6s^2~^2F_{7/2}$ electric octupole (E3) transition, are suitable for clock frequency measurements \cite{Taylor-PRA-1997, Roberts-PRA-1999, Webster-IEEE-2010, Huntemann-PRA-2014, Godun-PRL-2014, 
Huntemann-PRL-2014, Sanner-Nature-2019, Lange-PRL-2021, Filzinger-PRL-2023, Tofful-Metro-2024}. The clock frequency of the $4f^{14}6s~^2S_{1/2} 
\rightarrow 4f^{14}5d~^2D_{3/2}$ transition has been reported with an uncertainty of $3.3 \times 10^{-17}$ \cite{Lange-PRL-2021}. Two groups have
measured the clock frequencies of the $4f^{14}6s~^2S_{1/2} \rightarrow 4f^{13}6s^2~^2F_{7/2}$ transition with uncertainties of $2.7 \times 10^{-18}$ 
\cite{Sanner-Nature-2019} and $2.2 \times 10^{-18}$ \cite{Tofful-Metro-2024}. This suggests that a Yb$^+$ ion-based optical clock is one of the strong contenders to replace the caesium microwave clock as the primary frequency standard.

A particular advantage of the Yb$^+$ with multiple clock transitions is that it can be used to probe temporal and spatial variations of the 
fine-structure constant ($\alpha_e$) by measuring their clock frequency ratios with high precision \cite{Flambaum-CJP-2009}. With 
improved clock frequency measurements of the E2 and E3 transitions in Yb$^+$ over the years, limit on the inferred temporal variation of 
$\alpha_e$ has been gradually refined \cite{Godun-PRL-2014, Huntemann-PRL-2014, Lange-PRL-2021, Filzinger-PRL-2023}. A more stringent limit on the 
variation of $\alpha_e$ can be achieved by reducing uncertainties due to systematic effects in both the clock frequency and their ratio measurements.

The electric-field response properties, such as dipole polarizabilities and quadrupole moments, are essential for addressing and reducing systematic uncertainties in optical atomic clocks by accurately characterizing their sensitivity to external perturbations. Differential scalar polarizabilities quantify the response of clock states to external electric fields, including those arising from blackbody radiation (BBR). Tensor polarizabilities are also crucial for evaluating DC Stark shifts in clock experiments. Quadrupole shifts, resulting from interactions between the ion’s electric quadrupole moment and external electric field gradients, can be identified and reduced with precise calculations. This knowledge enables experimentalists to optimize trapping potentials and electric field gradients, reducing differential effects on clock states and minimizing systematic errors. The Stark and black-body radiation (BBR) shifts are more prominent in the $4f^{14}6s~^2S_{1/2} \rightarrow 4f^{14}5d~^2D_{3/2,5/2}$ transitions than the $4f^{14}6s~^2S_{1/2} \rightarrow 4f^{13}6s^2~^2F_{7/2}$ transition, which contributes significantly to the systematics of the E2 clock transitions in Yb$^+$. To calibrate these systematics and their uncertainties, accurate values of the electric-field response properties, such as the static electric dipole polarizabilities ($\alpha_d$), quadrupole moment ($\Theta$) and quadrupole polarizability ($\alpha_q$), of the $4f^{14}6s~^2S_{1/2}$ and $4f^{14}5d~^2D_{3/2}$ states are needed. In experiments, the differential value of $\alpha_d$ for the $4f^{14}6s~^2S_{1/2} \rightarrow 4f^{14}5d~^2D_{3/2}$ transition has been investigated by two groups. Though the final values in these experiments are within the quoted error bars, their central values differ significantly \cite{Schneider-PRL-2005, Baynham-2018}. A similar situation is observed for the measurement of the $\Theta$ value of the 
$4f^{14}5d~^2D_{3/2}$ state \cite{Schneider-PRL-2005, Lange-PRL-2020}.

Using different many-body methods, several calculations have been performed to obtain the $\alpha_d$ and $\alpha_q$ values for the 
$4f^{14}6s~^2S_{1/2}$ and $4f^{14}5d~^2D_{3/2}$ states in Yb$^+$ \cite{Migdalek-1982, Lea-2006, Safronova-PRA-2009, Porsev-PRA-2012, Roy-2017,
Chen-CPB-2023, Itano-PRA-2006, Latha-PRA-2007, Nandy-PRA-2014, Guo-CPB-2020}. However, these values show significant discrepancies compared to the measurements. For example, the differential $\alpha_d$ values for the $4f^{14}6s~^2S_{1/2} \rightarrow 4f^{14}5d~^2D_{3/2}$ clock 
transition in Yb$^+$ have been calculated using the pseudo-relativistic Hartree-Fock (HFR) method \cite{Lea-2006} and Fock-space relativistic 
coupled-cluster (RCC) theory \cite{Roy-2017}; neither of these results agree with the recently reported experimental value \cite{Baynham-2018}. 
The $\Theta$ value of the $4f^{14}5d~^2D_{3/2}$ state has been calculated using the multi-configuration Dirac-Hartree-Fock (MCDHF) method 
\cite{Itano-PRA-2006} and the RCC theory \cite{Latha-PRA-2007, Nandy-PRA-2014, Guo-CPB-2020}. Although the calculation using the RCC theory seems 
to match better with the experimental result than the MCDF result, the RCC value still differs by about 5\% from the measurement \cite{Lange-PRL-2020}. 
Other calculations use a hybrid of lower-order perturbation theory with another all-order method like configuration interaction (CI) or semi-empirical
methods. Such an approach may yield reasonably accurate results for specific properties but it is not reliable enough for high-accuracy calculations 
needed for the Yb$^+$ clock \cite{Migdalek-1982, Safronova-PRA-2009}. There are only a few experimental studies available for these quantities of the
$4f^{14}6s~^2S_{1/2} \rightarrow 4f^{14}5d~^2D_{3/2,5/2}$ clock transitions. Therefore, conducting more 
accurate calculations of the electric response properties of the Yb$^+$ clock transitions are imperative. In fact, experimental values for $\alpha_q$
of any of the states associated with both the E2 clock transitions are not yet known.

Since Yb$^+$ is a very heavy atomic system, electron correlation effects are quite strong in this ion. This makes it challenging to perform accurate 
calculations of atomic properties in Yb$^+$ \cite{Sahoo-PRA-2011, Nandy-PRA-2014}. The most accurate calculations of different properties of this ion 
reported previously were based on either the CI method or the Fock-space RCC method with singles and doubles approximation (RCCSD method) \cite{Sahoo-PRA-2011, 
Nandy-PRA-2014, Blythe-JPB-2003, Porsev-PRA-2012}. The truncated CI method suffers from a size-extensivity problem and is not suitable for application
to a heavy system like Yb$^+$. The property evaluation expression of the Fock-space RCC method does not satisfy the Hellman-Feynman theorem 
\cite{Bishop-TCA-1991}. In this work, we aim at presenting improved calculations of the $\alpha_d$, $\alpha_q$, and $\Theta$ values of the ground 
$4f^{14}6s~^2S_{1/2}$ state and metastable $4f^{14}5d~^2D_{3/2,5/2}$ states of Yb$^+$ using the general-order RCC method. Contributions from electron correlation 
effects are included through singles, doubles, and triples excitations of the RCC theory (RCCSDT method). Furthermore, the $\alpha_d$, $\alpha_q$,
and $\Theta$ values are extracted from the calculated energies using a finite-field (FF) approach. The advantage of using the FF approach is that 
it involves only the calculation of energies, which satisfies the Hellman-Feynman theorem in the RCC theory. In addition to providing accurate values of 
the $\alpha_d$, $\alpha_q$, and $\Theta$ values of the above states, we also intend to explain plausible reasons behind the significant discrepancies 
observed among the previously reported theoretical and experimental results. 

To carry out accurate calculations of atomic properties in Yb$^+$, a strategic plan is necessary, which we will discuss in more detail later. As a 
part of this plan, we consider the singly charged calcium ion (Ca$^+$) as a proxy system for Yb$^+$. The $4s~^2S_{1/2} \rightarrow 3d~^2D_{3/2,5/2}$ 
transition in Ca$^+$ are clock transitions. Accurate calculation of the $\alpha_d$ value for the $4s~^2S_{1/2}$ and $3d~^2D_{3/2, 5/2}$ states of the ion will help minimize uncertainties in the Ca$^+$ ion based optical clock. Thus, performing calculations in both Ca$^+$ and Yb$^+$ ions using the same methods can serve two purposes here. Since Ca$^+$ is a moderately heavy atomic system, it is possible to use a large set of basis functions while accounting for triple excitations more rigorously. From this analysis, it is possible to decide on an optimized basis set for performing calculations in Yb$^+$ to obtain accurate results for this ion. 

\section{Theory}

In the presence of an electric field of strength $|\mathcal{\vec{E}}|$, the energy level ($E_n$) of an atomic state can change due to the
interaction Hamiltonian $\vec{D} \cdot \mathcal{\vec{E}}$ for the electric dipole (E1) operator $\vec{D}$. Assuming the field is homogeneous and weak,
the modified energy level described by the angular momentum $J_n$ and its component $M_{J_n}$ can be expressed as
\begin{eqnarray}
\label{eq:alpha}
E_n^d(|\mathcal{\vec{E}}|) = E_n^{(d,0)} - \frac{1}{2} \alpha_d(J_n, M_{J_n}) |\mathcal{\vec{E}}|^2 - \cdots,
\end{eqnarray}
where $E_n^{(d,0)}$ is energy of the state in absence of the electric field. For weak fields of our interest, the above expression is approximated
up to the second term.

Similarly, for the field gradient $\vec{\nabla} |\mathcal{\vec{E}}|$ of a spatially varying electric field (assuming a zero electric field strength to suppress the quadratic Stark shift given above),
the energy level will be modified due to the interaction Hamiltonian $\vec{\Theta} \cdot \vec{\nabla} |\mathcal{\vec{E}}|$. If $\vec{\nabla}
|\mathcal{\vec{E}}|$ is weak, the modified energy level is expressed as
\begin{eqnarray}
\label{eq:E2}
E_n^q(\vec{\nabla} |\mathcal{\vec{E}}|) &=& E_n^{(q,0)} - \frac{1}{2} \Theta(J_n, M_{J_n}) \vec{\nabla} |\mathcal{\vec{E}}|
\nonumber \\ && - \frac{1}{8} \alpha_q (J_n, M_{J_n}) (\vec{\nabla} |\mathcal{\vec{E}}|)^2 - \cdots.
\end{eqnarray}
In this case, we approximate the expression up to the third-term.

For convenience, $\alpha_d(J_n, M_{J_n})$ and $\alpha_q (J_n, M_{J_n})$ in Eqs. (\ref{eq:alpha}) and (\ref{eq:E2}) are written as
\begin{eqnarray}
\alpha_{d,q}(J_n, M_{J_n}) &=& \alpha_{d,q}^S + \frac{3M_{J_n}^2 - J_n(J_n + 1)}{J_n(2J_n - 1)} \alpha_{d,q}^T,
\end{eqnarray}
where $\alpha_{d,q}^S$ and $\alpha_{d,q}^T$ are referred to as the scalar and tensor polarizabilities, respectively, and given by
\begin{eqnarray}
\alpha_{d,q}^S &=& \frac{1}{2J_n + 1} \sum_{M_{J_n}} \alpha_{d,q}(J_n, M_{J_n})
\label{eqs}
\end{eqnarray}
and
\begin{eqnarray}
\alpha_{d,q}^T &=& \alpha_{d,q}(J_n, |M_{J_n}| = J_n) - \alpha_{d,q}^S.
\label{eqt}
\end{eqnarray}

In Eq. (\ref{eq:E2}), $\Theta(J_n, M_{J_n})$ gives the electric quadrupole moment for the corresponding $J_n$ and $M_{J_n}$ level of a state.
As a practice, $\Theta$ is estimated for the $|M_{J_n}| = J_n$ level and its value for another $M_{J_n}$ level is estimated by multiplying
with the corresponding angular factors. All the physical quantities are given in atomic units (a.u.) unless specified otherwise.

\begin{table*}[btp]
	\caption{Static scalar ($\alpha_d^S$ ) and tensor ($\alpha_d^T$) electric dipole polarizabilities (in a.u.) of the $4s~^2S_{1/2}$ and the $3d~^2D_{5/2,3/2}$ states of Ca$^+$, and the differential scalar electric dipole polarizabilities ($\Delta \alpha_d^S$) with respect to the ground state.  \label{tab:Ca+}}
	{\setlength{\tabcolsep}{2pt}
\begin{tabular}{lcc ccc ccc cc }\hline\hline			\addlinespace[0.2cm]
\multirow{2}{*}{Methods}	&&	$4s~^2S_{1/2}$	&&	\multicolumn{3}{c}{$3d~^2D_{5/2}$}					&&	\multicolumn{3}{c}{$3d~^2D_{3/2}$}					\\ \cline{5-7}  \cline{9-11} \addlinespace[0.1cm]
&&	$\alpha_d^S$	&&	$\alpha_d^S$ 	&	$\Delta \alpha_d^S$	&	$\alpha_d^T$ 	&&	$\alpha_d^S$	&	$\Delta \alpha_d^S$	&	$\alpha_d^T$ 	\\ \hline\addlinespace[0.2cm]
e9-SD $<$ 10	&&	75.975 	&&	32.657 	&	-43.318 	&	-24.086 	&&	38.820 	&	-37.155 	&	-15.300 	\\
e9-SDT $<$ 10	&&	74.449 	&&	31.243 	&	-43.206 	&	-25.193 	&&	34.662 	&	-39.788 	&	-20.162 	\\
$\Delta_{P_T}$-e9	&&	-1.525 	&&	-1.414 	&	0.112 	&	-1.107 	&&	-4.158 	&	-2.633 	&	-4.862 	\\\addlinespace[0.1cm]
e15-SD $<$ 10	&&	75.726 	&&	32.074 	&	-43.652 	&	-23.540 	&&	35.685 	&	-40.041 	&	-13.843 	\\
e15-SDT $<$ 10	&&	74.147 	&&	31.046 	&	-43.101 	&	-25.018 	&&	33.676 	&	-40.471 	&	-17.133 	\\
$\Delta_{P_T}$-e15	&&	-1.579 	&&	-1.028 	&	0.550 	&	-1.478 	&&	-2.009 	&	-0.430 	&	-3.290 	\\\addlinespace[0.1cm]
e17-SD $<$ 10	&&	75.931 	&&	32.029 	&	-43.901 	&	-23.519 	&&	35.440 	&	-40.491 	&	-14.100 	\\
e17-SD $<$ 50	&&	75.969 	&&	31.912 	&	-44.057 	&	-23.342 	&&	35.261 	&	-40.708 	&	-13.815 	\\
e17-SD $<$ 100	&&	76.065 	&&	32.024 	&	-44.041 	&	-23.430 	&&	35.381 	&	-40.684 	&	-13.871 	\\\addlinespace[0.1cm]
e19-SD $<$ 100	&&	76.140 	&&	32.019 	&	-44.121 	&	-23.423 	&&	35.373 	&	-40.767 	&	-13.879 	\\
\\\addlinespace[0.2cm]
Final	&&	74.62(41)	&&	30.59(6)	&	$-44.02(47)$ 	&	$-24.50(12)$	&&	33.36(31)	&	$-41.25(72)$ 	& $ -17.17(10)$	\\ \hline \addlinespace[0.2cm]
CI+All-order-2007 \cite{Bindiya-PRA-2007}	&&	76.1(1.1)	&&	32.0(1.1)	&	-44.1	&	-24.5(4)	&&		&		&		\\
RCC-2009 \cite{Sahoo-PRA-2009}	&&	73.0(1.5)	&&	29.5(1.0)	&	-43.5	&	-22.45(5)	&&	28.5(1.0)	&	-44.5	&	-15.8(7)	\\
Exp.-2019 \cite{Huang-PRA-2019}	&&		&&		&	-44.07(1)	&		&&		&		&		\\
NR-SOO-2009 \cite{Theodosiou-PRA-1995}	&&		&&		&		&		&&	32.73	&		&	-25	\\
NR-SOO-1998 \cite{Barklem-PRA-1998}	&&		&&		&		&		&&	25.4	&		&		\\ \hline\hline
\end{tabular}}
\end{table*}

\section{Method of calculation}

We consider the approximated Dirac-Coulomb-Gaunt (DCG) Hamiltonian in the present work to calculate the atomic wave functions and energies, given in
a.u. by
\begin{eqnarray}
	\hat{H}&=&\sum_i [c(\vec { \bm{\alpha}}\cdot \vec {\bf p})_i+(\bm \beta-1)_i c^2+V_{nuc}(r_i) ]  \nonumber \\
	&& + \sum_{i<j}\bigg[\frac{1}{r_{ij}}-\frac{1}{2}\frac{ {\vec {\bm \alpha}}_i \cdot {\vec { \bm \alpha}}_j }{r_{ij}}\bigg], \label{DCG}
\end{eqnarray}
where $ \vec {\bm \alpha}$ and  $\bm \beta$ are the Dirac matrices, $\vec {\bf p}$ is the momentum operator, $c$ denotes the speed of light, and
$V_{nuc}(r_i)$ is the nuclear potential. Here we have subtracted the rest mass energies of electrons and use Gaussian charge distribution to
define the nuclear potential.

We obtain first the mean-field wave function using the Dirac-Hartree-Fock (DHF) method with the DCG Hamiltonian from the Dirac package \cite{Dirac,Dirac2020}. In the DHF method, we have used the uncontracted Gaussian-type consistent basis sets. The kinetic balance condition has been imposed between the large and small components of the DHF orbitals.

In the RCC theory method, the exact wave function of the ground state is expressed as
\begin{equation}
|\Psi_0 \rangle = e^{\hat{T}}|\Phi_0 \rangle,\label{CCwave}
\end{equation}
where $\hat{T}$ is the RCC excitation operator that constructs all possible excitation configurations acting upon the DHF wave function
$|\Phi_0 \rangle$. In the general-order RCC method \cite{Kallay-JCP-2002,Kallay-JCP-2004}, the $T$ operator can be given using the second-quantization
operators as
\begin{eqnarray}
\hat{T} &=& \hat{T}_1 + \hat{T}_2 + \cdots \nonumber \\
     &=& \sum_{i,a} t_i^a a_a^{\dagger} a_i + \sum_{i<j,a<b} t_{ij}^{ab} a_a^{\dagger} a_b^{\dagger} a_j a_i + \cdots.\label{CCampl}
\end{eqnarray}
Here $\hat{T}_K$ means a $K^{th}$ level excitation that is responsible for exciting electrons from the $i,j,k,\cdots$ occupied orbitals to the
$a,b,c,\cdots$ virtual orbitals with the amplitudes $t_{i,j,k,\cdots}^{a,b,c,\cdots}$. The ground state energy ($E_0$) and amplitudes of the
$\hat{T}$ operator can be obtained by solving equations
\begin{equation}
\langle \Phi_0|e^{-\hat{T}}\hat{H}e^{\hat{T}}|\Phi_0 \rangle=E_0,
\label{eq1}
\end{equation}
and
\begin{equation}
\langle\Phi_K|e^{-\hat{T}} \hat{H} e^{\hat{T}}| \Phi_0\rangle=0 ,
\label{eq2}
\end{equation}
where $|\Phi_K \rangle = \left \{ a_a^{\dagger} a_b^{\dagger} a_c^{\dagger} \cdots a_k a_j a_i \right \}_K |\Phi_0 \rangle$ is the $K^{th}$ level
excitation Slater determinant with respect to $|\Phi_0\rangle$. In the singles and doubles excitation approximated RCC theory (RCCSD method),
$\hat{T}=\hat{T}_1 + \hat{T}_2$ and in the RCCSDT method approximation $\hat{T}=\hat{T}_1 + \hat{T}_2 + \hat{T}_3$.

In the next step, we define the exact $L^{th}$ level excited state of the considered atomic system by
\begin{eqnarray}
 |\Psi_L \rangle &=& R_L |\Psi_0 \rangle
     = R_L  e^{\hat{T}}|\Phi_0 \rangle
     =  e^{\hat{T}} R_L |\Phi_0 \rangle ,
\end{eqnarray}
where $R_L$ is another excitation operator like $\hat{T}$ but it acts on the exact ground state wave function $|\Psi_0 \rangle$. Thus in the
second-quantization notation, it will have similar form like $\hat{T}$ but the excitation amplitudes of $R_L$ will be different than $\hat{T}$. These amplitudes are
obtained by the following equation
\begin{equation}
\langle \Phi_K|e^{-\hat{T}} \hat{H} e^{\hat{T}} R_L |\Phi_0 \rangle = E_L \delta_{L,K} \langle \Phi_K| R_L |\Phi_0 \rangle ,
\label{eq3}
\end{equation}
where $E_L$ is the energy eigenvalue of the desired excited state of the considered system. For all practical purposes, this equation is given using
the commutation relation
\begin{eqnarray}
\langle \Phi_K| [e^{-\hat{T}} \hat{H} e^{\hat{T}}, R_L] |\Phi_0 \rangle &=& (E_L - E_0) \delta_{L,K} \langle \Phi_K| R_L |\Phi_0 \rangle   \nonumber \\
    &=&  \Delta E_L \delta_{L,K} \langle \Phi_K| R_L |\Phi_0 \rangle ,
\end{eqnarray}
where $\Delta E_L$ is the excitation energy of the $L^{th}$ level. The advantage of using this equation is that it gives the excitation energies of
the desired states directly without computing the ground state energy $E_0$. Moreover, the amount of computation in this method can be reduced by
using a normal order Hamiltonian. As can be noticed the above equation gives a non-hermitian matrix, which is a challenge to store in a large
configuration dimensional space for the matrix diagonalization. Thus, we first carry out calculations using the RCCSD method approximation with a
large set of basis functions. Following which calculations are performed using both the RCCSD and RCCSDT methods with a reasonable size of basis
functions. To obtain accurate results in the end, we use the RCCSD results from the largest basis as the base then the differences between the results
from the RCCSDT and RCCSD methods from a smaller basis are added to the RCCSD results of the larger basis. We generated all the required one-body and 
two-body integrals using the DIRAC program package \cite{Dirac,Dirac2020} for carrying out all these calculations. The excited states are obtained
using the MRCC package \cite{MRCC}, which is interfaced with the DIRAC program.

To estimate the $\alpha_d(J_n,M_{J_n})$ values, we adopt the FF approach in which an effective atomic Hamiltonian,
\begin{eqnarray}
\hat{H}_{eff}^d = \hat{H} + \vec D \cdot \mathcal{\vec E} ,
\label{heffd}
\end{eqnarray}
is used in the RCC methods after obtaining the unperturbed solutions. We performed a series of calculations at the RCCSD and RCCSDT methods by varying
electric field strength $|\mathcal{\vec E}|$, then the obtained energies are fitted into the Taylor expansion at $|\mathcal{\vec E}| \equiv 0$. The
coefficient of the $2^{nd}$ term in the Taylor's expansion corresponds to the $\alpha_d(J_n,M_{J_n})$ value of the respective state determined using
the RCCSD and RCCSDT methods. Further, the $\alpha_d^S$ and $\alpha_d^T$ values are extracted using Eqs. (\ref{eqs}) and (\ref{eqt}), respectively,
after calculating the $\alpha_d(J_n,M_{J_n})$ values for all possible $M_{J_n}$ values. The $\Theta$ and $\alpha_q$ values are also evaluated through
a similar procedure by using the effective Hamiltonian
\begin{eqnarray}
\hat{H}_{eff}^q = \hat{H} + \vec \Theta \cdot \vec \nabla |\mathcal{\vec E}|.
\label{heffq}
\end{eqnarray}
From the polynomial fitting of energies using the above Hamiltonian with different $\vec \nabla |\mathcal{\vec E}|$ values, the second and third terms give us the
$\Theta(J_n,M_{J_n})$ and $\alpha_q(J_n,M_{J_n})$ values, respectively. In this work, we have varied $|\mathcal{\vec E}|$ between $[0-0.0025]$ a.u. to estimate
the $\alpha_d(J_n,M_{J_n})$ values and the $\vec \nabla |\mathcal{\vec E}|$ values are varied between $[1.5-30]\times 10^{-6}$ a.u. to infer
the $\Theta(J_n,M_{J_n})$ and $\alpha_q(J_n,M_{J_n})$ values. Thresholds are set to be $10^{-10}$ and $10^{-6}$ for the ground and excited states,
respectively, in order to achieve convergence in the calculated energies using the respective effective Hamiltonians.

\section{Results and Discussion}

\begin{table*}[btp]
	\caption{Calculated total ground energies ($E_0$) of Yb$^{+}$ and Yb$^{2+}$ and their differences as ionization potential, i.e. $\Delta E_0$. These values are given in cm$^{-1}$. The extracted ground state electric dipole and quadrupole polarizabilities of these ions are also given 
	in a.u. using the FF approach. Uncertainties are quoted within the parenthesis. \label{tab:Yb+2S}}
	{\setlength{\tabcolsep}{6pt}
		\begin{tabular}{lccccccc}\hline\hline			\addlinespace[0.2cm]
Methods	&	$E_0$ (Yb$^+$)	&	$E_0$ (Yb$^{2+}$)	&	$\Delta E_0$	&	$\alpha_d^S$ (Yb$^{+}$) 	&	$\alpha_q^S$ (Yb$^{+}$) 	&	$\alpha_d^S$ (Yb$^{2+}$)	\\ \hline\addlinespace[0.2cm]
$2\zeta$	&		&		&		&		&		&		\\
e23-SD  $<$ 10	&	$-14052.3714$ 	&	$-14051.9303$ 	&	96818 	&	65.75 	&	709.68 	&	7.24 	\\
e23-SDT  $<$ 10	&	$-14052.4163$ 	&	$-14051.9716$ 	&	97610 	&	63.48 	&	686.16 	&	7.68 	\\
${\Delta_{P_T}}_{-2\zeta}$	&	$-0.0449$ 	&	$-0.0413$ 	&	792 	&	$-2.27$ 	&	$-23.52$ 	&	0.44 	\\\addlinespace[0.2cm]
$3\zeta$ 	&		&		&		&		&		&		\\
e23-SD $<$ 10	&	$-14052.6231$ 	&	$-14052.1798$ 	&	97203 	&	64.40 	&	757.52 	&	7.23 	\\
e33-SD $<$ 10	&	$-14052.9033$ 	&	$-14052.4592$ 	&	97471 	&	64.11 	&	745.73 	&	7.21	\\
e23-SDT $<$ 10 	&	$-14052.6674$ 	&	$-14052.2203$ 	&	98126 	&	63.87 	&		&	7.65 	\\
e23-SD $<$ 50	&	$-14052.9097$ 	&	$-14052.4665$ 	&	97267 	&	64.49 	&	773.26 	&	7.25 	\\
e23-SD $<$ 10+saug	&	$-14052.6238$ 	&	$-14052.1802$ 	&	97360 	&	64.44 	&	771.76 	&	7.22 	\\
${\Delta_{P_T}}_{-3\zeta}$	&	$-0.0442$ 	&	$-0.0405$ 	&	923 	&	$-0.53$ 	&		&	0.43 	\\
$\Delta_{4d}$	&	$-0.2802$ 	&	$-0.2794$ 	&	268 	&	$-0.28$ 	&	$-11.79$ 	&	$-0.02$ 	\\
$\Delta_{virt}$	&	$-0.2866$ 	&	$-0.2867$ 	&	64 	&	0.10 	&	15.74 	&	0.02 	\\
$\Delta_{saug}$	&	$-0.0007$ 	&	$-0.0004$ 	&	157 	&	0.04 	&	14.24 	&	$-0.01$ 	\\\addlinespace[0.2cm]
$4\zeta$ 	&		&		&		&		&		&		\\
e23-SD $<$ 10	&	$-14052.6143$ 	&	$-14052.1707$ 	&	97364 	&	64.23 	&	778.16 	&	7.22 	\\
$\Delta_{P_T}$	&	$-0.0442$ 	&	$-0.0405$ 	&	923 	&	$-0.53$ 	&	$-23.52$ 	&	0.43 	\\
$\Delta_{basis}$	&	0.0088 	&	0.0091 	&	161 	&	$-0.17$ 	&	20.64 	&	$-0.01$ 	\\
\addlinespace[0.3cm]
Final 	&	$-$14052.65(40)	&	$-$14052.2(4)	&	98447(355)	&	63.53(35)	&	775(32)	&	7.64(3)	\\
\hline
\addlinespace[0.2cm]
NIST \cite{NIST}	&		&		&	98231.75	&		&		&		\\
RHF-2006  \cite{Lea-2006} 	&		&		&		&	58.4	&		&		\\
RMP+CP(Ia)-1982 \cite{Migdalek-1982}	&		&		&		&	62.84	&		&	7.36	\\
RMP+CP(Ib)-1982 \cite{Migdalek-1982}	&		&		&		&	59.32	&		&	11.14	\\
RMBPT+RPA-2009  \cite{Safronova-PRA-2009}	&		&		&		&	62.04	&		&	6.386	\\
RCCSD-2017 \cite{Roy-2017}	&		&		&		&	59.3(8)	&		&	7.72	\\
RCICP-2023  \cite{Chen-CPB-2023}	&		&		&		&	60.1(3.1)	&		&		\\\hline\hline
	\end{tabular}}
\end{table*}

In Table \ref{tab:Ca+}, we present results for the $4s~^2S_{1/2} \rightarrow 3d~^2D_{3/2,5/2}$ clock transitions of Ca$^+$ calculated using both the RCCSD and RCCSDT 
methods. The Ca$^{+}$ ion is relatively light, which enables us to consider correlations among all the electrons and use the largest relativistic 
basis set, dyall.cv4z \cite{Dyall-Ca}, with $d$ functions, in the calculations for this ion. To ensure reliability in the calculated values of $\alpha_d$, we 
performed a detailed analysis of how the results depend on various approximations in the computational procedure adopted in this work. For this 
purpose, we allowed correlations among different numbers of occupied electrons gradually. Our `e9', `e17', and `e19' notations are used here to denote 
approximations representing 9 ($3s^23p^64s$), 17 ($2s^22p^63s^23p^64s$), and 19 ($1s^22s^22p^63s^23p^64s$) occupied electrons that are allowed for 
correlations in the RCCSD method, labeled as  `SD' in the result table, and RCCSDT method, labeled as `SDT' in the result table. We also tested the convergence of the results with an 
increasing number of virtual spinors by relaxing the upper-level energy cut-off. Notations `$<$10', `$<$50', and `$<$100' used in the approximations 
denote energy cut-off for virtuals as 10 a.u., 50 a.u., and 100 a.u., respectively.

As shown in Table \ref{tab:Ca+}, the $\alpha_d^S$ and $\alpha_d^T$ values for the $4s~^2S_{1/2}$ and $3d~^2D_{5/2}$ states in Ca$^+$ exhibit excellent 
convergence as more active occupied electrons and virtual spinors are included in the calculations. Additionally, we estimated contributions from the 
triple excitations, denoted by $\Delta P_T$, by comparing the differential results between the e9-SD$<$10 and e9-SDT$<$10 approximations (denoted as 
`$\Delta_{P_T}$-e9'), as well as between the e15-SD$<$10 and e15-SDT$<$10 approximations (denoted as `$\Delta_{P_T}$-e15'). These values are almost agreeing
with each other suggesting that triple effects from high-lying orbitals may not be that significant. Consequently, the final results are taken from the e19-SD$<$100 calculations, after adding the $\Delta_{P_T}$-e15 contributions. The 
absolute differences between the e15-SD$<$10 and e19-SD$<$100 results are used to assign uncertainties to the final values.

Our recommended final values for $\alpha_d^S$ are 74.62(41) a.u. for the $4s~^2S_{1/2}$ state and 30.59(6) a.u. for the $3d~^2D_{5/2}$ state of 
Ca$^+$. For the $3d~^2D_{5/2}$ state, the $\alpha_d^T$ value is determined to be $-24.50(12)$ a.u.. For the scalar polarizability, we observe a discrepancy
of approximately 2.0 a.u. for both the $4s~^2S_{1/2}$ and $3d~^2D_{5/2}$ states when comparing our values to the results obtained using the 
CI+All-order method in 2007, denoted as `CI+All-order-2007' \cite{Bindiya-PRA-2007}, and the RCCSD method in 2009, denoted as `RCCSD-2009'
\cite{Sahoo-PRA-2009}. However, these discrepancies reduce once $\Delta P_T$ contributions are included in our calculations. The $\Delta P_T$ contributions account for approximately 2-3\% in $\alpha_d$ for the $4s~^2S_{1/2}$ and $3d~^2D_{5/2}$ states and about 1\% for their $\Delta \alpha_d$ value. Although these contributions are small, they play a crucial role in achieving better consistency with the experimental results. This highlights the importance of incorporating electron correlations from triple excitations for accurately estimating the differential scalar polarizability 
($\Delta \alpha_d^S$) for the $4s~^2S_{1/2} \rightarrow 3d~^2D_{5/2}$ clock transition. In Ref. \cite{Bindiya-PRA-2007}, a sum-over-states approach (referred to 
as the sum-over method) was employed, in which main contributions were derived by combining experimental energy values with the calculated E1 
matrix elements among the low-lying states. The tail contributions from the high-lying states were estimated using the DHF method, and core 
contributions were calculated via the random phase approximation (RPA). In contrast, the calculations in Ref. \cite{Sahoo-PRA-2009} were performed 
using a linear response approach within the RCCSD approximation. Our final recommended value for $\Delta \alpha_d^S$ of the Ca$^+$ clock transition 
is $-44.02(47)$ a.u., which shows excellent agreement with both the previous theoretical results \cite{Bindiya-PRA-2007,Sahoo-PRA-2009} and experimental
data \cite{Huang-PRA-2019}. Similarly, our estimated $\alpha_d^T$ value is in good agreement with previous calculations. For the tensor polarizability,
the inclusion of the triple excitation contributions significantly affects the $\alpha_d^T$ value for the $3d~^2D_{5/2}$ state, bringing it closer to 
the `CI+All-order-2007' result of Ref. \cite{Bindiya-PRA-2007}. For the $3d~^2D_{3/2}$ state, our calculation yields $\alpha_d^S = 33.36(31)$ a.u. and $\alpha_d^T = -17.17(10)$ a.u., which are comparatively larger than values reported in previous studies. It is important to note that this work represents the first application of the FF approach in RCCSD and RCCSDT approximations. Due to implementation differences, we are unable to reproduce some of the previously reported results, such as the RCCSD calculation by Sahoo et al. \cite{Sahoo-PRA-2009}. However, our value for $\alpha^S_d$ is consistent with the non-relativistic sum-over-oscillation (NR-SOO) results by Theodosiou \cite{Theodosiou-PRA-1995}, which provides some validation for our approach. This highlights the computational challenges involved in accurately determining electric response properties, particularly for the excited states. We anticipate that future investigations, incorporating alternative methods and more optimized numerical basis sets, will help resolve these discrepancies.

After recognizing significance of electron correlation effects in the accurate determination of the E1 polarizabilities of Ca$^+$, we now extend
our investigation to explore the roles of correlation effects in evaluating the polarizabilities of the Yb$^+$ ion. Additionally, we aim at estimating 
the $\alpha_d^S$ value of the ground state of Yb$^{2+}$ accurately using the same basis sets employed for calculating the wave functions of Yb$^+$.  
This, in principle, can serve two purposes: it can explain the large discrepancies among previously reported Yb$^{2+}$ results from various 
calculations and then, it can help validate our calculations for Yb$^+$, as both the methods and basis functions used for Yb$^+$ and Yb$^{2+}$ are the same for calculating 
the ground state polarizabilities. The results for the ground-state polarizabilities of Yb$^+$ and Yb$^{2+}$ are summarized in Table \ref{tab:Yb+2S}.

We performed RCCSD and RCCSDT calculations twice: first, to determine the ground state energy ($E_0$) of Yb$^{2+}$, and second, to calculate the 
energies of Yb$^+$. Since Yb$^{2+}$ is a closed-shell atomic system, we used Eq. (\ref{eq1}) to compute its ground state energy. From the ground state
energy difference between both ions ($\Delta E_0 = E_0(\text{Yb}^+) - E_0(\text{Yb}^{2+})$), we obtained the electron affinity (negative of the 
ionization potential) of Yb$^+$. The ground state $\alpha_d^S$ values for both the Yb$^+$ and Yb$^{2+}$ ions, along with the $\alpha_q^S$ value for 
Yb$^+$, are inferred from these energy values. We use the notations `e23' and `e33' to denote approximations involving 23 ($5s^25p^64f^{14}6s$) and 
33 ($4d^{10}5s^25p^64f^{14}6s$) occupied electrons, which are considered in the RCCSD and RCCSDT correlation calculations, respectively. The `$<$10' 
and `$<$50' notations refer to the truncation of virtual orbitals with energies less than 10 a.u. and 50 a.u., respectively. All calculations are 
performed using the dyall.cv2z, dyall.cv3z, and dyall.cv4z basis sets \cite{Gomes-TCA-2010}, labeled $2\zeta$, $3\zeta$, and $4\zeta$, respectively. 
To compensate for the absence of diffusion functions in these basis sets, we extended the dyall.cv3z basis set by adding a single-fold diffusion 
function using an even-temper expansion, denoted as `saug'.

Our results are based on the RCCSD calculations with 23 correlated electrons, using a truncation of virtual orbitals below 10 a.u., denoted as 
`e23-SD$<$10' for the dyall.cv4z basis set. Contributions from the triple excitations through the RCCSDT method is accounted with relatively smaller 
basis sets, namely $2\zeta$ and $3\zeta$, compared to the RCCSD calculations. The difference between the e23-SD$<$10 and e23-SDT$<$10 results provides
an estimate of the triple excitation contributions, $\Delta P_T$. Consequently, the final values are obtained from the e23-SD$<$10 results using the 
$4\zeta$ basis set, with the $\Delta P_T$ correction estimated from the $3\zeta$ calculations. In Table \ref{tab:Yb+2S}, we also provide the value of $\alpha_q^S$ for the Yb$^+$ ground state and $\alpha_d^S$ for the Yb$^{2+}$ core. The $\Delta_{P_T}$ contributions are approximately 1\% for $\alpha_d^S$ (Yb$^+$), 3\% for $\alpha_q^S$ (Yb$^+$), and 6\% for $\alpha_d^S$ (Yb$^{2+}$). Though these contributions are relatively small, they can affect while estimating systematic effects in the high-precision measurements.

Uncertainties in our results are estimated from various sources. First, we assessed the influence of inner-core electron excitations on the calculated
values by including the $4d$ core electrons in the RCCSD calculations. By comparing results from the e23-SD$<$10 and e33-SD$<$10 approximations, we 
estimated the core electron contribution, denoted as $\Delta_{4d}$, as reported in the same table. Additionally, the impact of high-lying virtual 
orbitals were determined by comparing the results from the e23-SD$<$10 and e23-SD$<$50 calculations, labeled as $\Delta_{virt}$. We also examined the 
effect of the diffusion-augmented basis set by comparing the e23-SD$<$10 and e23-SD$<$10 saug results, denoted as $\Delta_{saug}$, which provides 
insight into the influence of the augmented basis set. To account for errors related to the finite basis set size, denoted as $\Delta_{basis}$, we 
estimated this by taking the difference between the e23-SD$<$10 results obtained with the $3\zeta$ and $4\zeta$ basis sets. All these contributions to 
the uncertainties from $\Delta_{4d}$, $\Delta_{virt}$, $\Delta_{saug}$, and $\Delta_{basis}$ are combined in quadrature to yield the net
uncertainty in our final results. They are quoted within the parentheses for the final values in each table.

The total energies $E_0$ of the ground states of Yb$^+$ and Yb$^{2+}$ cannot be directly measured in experiments, so we compare our calculated 
$\Delta E_0$ value with the experimental ionization potential of the Yb$^+$ ion \cite{NIST}. The excellent agreement between our calculations and the 
experimental value supports the reliability of our method. When comparing our calculated ground state $\alpha_d^S$ value with previously reported 
theoretical results, we find that earlier estimated values are significantly lower than ours. For example, the value obtained using the approximated 
relativistic Hartree-Fock method (RHF-2006) is $\alpha_d^S = 58.4$ a.u. \cite{Lea-2006}, while the relativistic model potential combined with the 
core-potential (RMP+CP) method yields a value of $\alpha_d^S = 59.32$ a.u., when the core potential is defined using $\alpha_d = 11.14$ a.u. for 
Yb$^{2+}$ (RMP+CP(Ia)-1982). However, when the core potential is defined using $\alpha_d^S = 7.36$ a.u. for Yb$^{2+}$, the result is 
$\alpha_d^S = 62.84$ a.u. (RMP+CP(Ib)-1982) \cite{Migdalek-1982}. Our results indicate $\alpha_d^S$ for Yb$^{2+}$ as 7.64(3) a.u., calculated using the 
RCCSD and RCCSDT methods, suggesting that the latter result of $\alpha_d^S = 62.84$ a.u. (RMP+CP(Ib)-1982) is more reasonable. The sum-over-states 
approach, using E1 matrix elements from the RCCSD method, yields $\alpha_d^S = 59.3(8)$ a.u. for the ground state of Yb$^+$ (RCCSD-2017) 
\cite{Roy-2017}. Another calculation using relativistic many-body perturbation theory combined with RPA gives a value of $\alpha_d^S = 62.04$ a.u. 
(RMBPT+RPA-2009) \cite{Safronova-PRA-2009}, which is closer to our result but is still on slightly lower-side. The $\alpha_d^S$ values obtained by 
the relativistic configuration interaction plus core polarization (RCICP) method, reported in 2023 (RCICP-2023), are obviously lower than our result, 
but is found to be within the error bar. These comparisons suggest that previous calculations of the ground state E1 polarizability of Yb$^+$ 
have consistently underestimated the value compared to our RCCSD and RCCSDT results obtained using the MRCC program. Therefore, experimental 
confirmation of these results is necessary.

\begin{table}[btp]
	\caption{Excited energies (EE, cm$^{-1}$), static scalar ($\alpha_d^{S}$) and tensor ($\alpha_d^{T}$) electric dipole polarizabilities (in a.u.) of the $4f^{14}5d~^2D_{3/2}$ state, and the differential $\Delta \alpha_d^S$ value of the $4f^{14}6s~^2S_{1/2} \rightarrow
	4f^{14}5d~^2D_{3/2}$ clock transition in Yb$^+$. \label{tab:Yb+2D23}}
	{\setlength{\tabcolsep}{1pt}
		\begin{tabular}{lcc cc}\hline\hline			\addlinespace[0.2cm]
Methods	&	EE	&	 $\alpha_d^S$ 	&	$\alpha_d^T$	&$\Delta \alpha_d^S$	\\ \hline\addlinespace[0.2cm]
$2\zeta$	&		&		&		&		\\
e23-CCSD $<$ 10	&	23313 	&	119.71 	&	$-81.37$ 	&	56.18 	\\
e23-CCSDT $<$ 10	&	23126 	&	103.31 	&	$-74.63$ 	&	39.78 	\\ \addlinespace[0.2cm]
$3\zeta$	&		&		&		&		\\
e23-CCSD $<$ 10	&	23619 	&	110.46 	&	$-73.54$ 	&	46.93 	\\
e23-CCSD $<$ 10 saug	&	23595 	&	111.58 	&	$-74.66$ 	&	48.05 	\\ \addlinespace[0.2cm]
$4\zeta$	&		&		&		&		\\
e23-CCSD $<$ 10	&	23558 	&	108.36 	&	$-70.98$ 	&	44.83 	\\
$\Delta_{basis}$	&	$-61$ 	&	$-2.10$ 	&	2.56 	&	$-2.10$ 	\\ 
\addlinespace[0.2cm]
Final 	&	23065(123)	&	101(4)	&	$-$72(5)	& 38(4)	\\
\hline\addlinespace[0.2cm]
NIST \cite{NIST}	&	22960.80 	&		&		&		\\
Exp.-2005  \cite{Schneider-PRL-2005}	&		&		&	$-82(13)$	&	41.8(8.5)	\\
RHF  \cite{Lea-2006} 	&		&	89.93	&	$-73.56$	&	31.53	\\
RCC-2017   \cite{Roy-2017}	&		&	107(3) 	&	$-75.2$ 	&	47.7(3.8)	\\
Exp.-2018   \cite{Baynham-2018} 	&		&		&	$-72.0(2.3)$ 	&	35.7(1.8)	\\
RCICP-2023  \cite{Chen-CPB-2023}	&		&99.1(5.2)	&	$-69.3(3.6)$	&		\\ \hline\hline
	\end{tabular}}
\end{table}

Table \ref{tab:Yb+2D23} summarizes the results for the  $^2D_{3/2}$ state of Yb$^+$ obtained using the RCCSD and RCCSDT methods using the 2$\zeta$, 3$\zeta$, and 4$\zeta$ basis sets. The excited energy (EE) values show good convergence and align well with the experimental values \cite{NIST}. However, the convergences for $\alpha_d^S$ and $\alpha_d^T$ values are more challenging. From the calculations with the 2$\zeta$ basis set, we observe that the $\alpha_d^S$ values are relatively large in the e23-SD$<$10 approximation. When contributions from triple excitations (SDT) are included, the value decreases, suggesting that higher-order correlation effects are contributing with opposite signs. This reduction emphasizes the importance of considering higher-order correlations to achieve more accurate polarizability values. Further, calculations using the $3\zeta$ and $4\zeta$ basis sets, which are larger, suggest that the $\alpha_d^S$ and $\alpha_d^T$ results can be refined further. Adding diffuse functions (saug) to the 3$\zeta$ basis set leads to slight variations in the obtained values, indicating that the inclusion of augmented basis functions does not significantly affect the final results.

Although the RCCSDT calculations with the 3$\zeta$ basis set can provide more accurate results, they are computationally expensive. Therefore, we base our final values on the RCCSDT results with the 2$\zeta$ basis set, and we estimate the basis set correction ($\Delta_{basis}$) using the differences in the e23-SD$<$10 results from the 3$\zeta$ and 4$\zeta$ basis sets. After applying the $\Delta_{basis}$ correction to the 2$\zeta$-based e23-SDT$<$10 results, we obtain the final values: $\alpha_d^S = 101(4)$ a.u. and $\alpha_d^T =-72(5)$ a.u.
This leads to a differential $\Delta \alpha_d^S$ for the $^2S_{1/2} \rightarrow ^2D_{3/2}$ clock transition in Yb$^+$ of 38(4) a.u., where the ground state polarizability is considered as $\alpha^S_d = 65.53(35)$. This result is lower than the experimental value from Exp.-2005 \cite{Schneider-PRL-2005} and the RCCSD result from RCCSD-2017 \cite{Roy-2017}. However, it aligns more closely with the experimental result from Exp.-2018 \cite{Baynham-2018}. The result from the RHF method (RHF-2006) \cite{Lea-2006} appears to underestimate the $\Delta \alpha_d^S$ value. For the tensor polarizability $\alpha_d^T$, our result is consistent with the earlier experimental result from 2015 (Exp-2015) \cite{Schneider-PRL-2005} within the error bar and more closely with the latter experimental result (Exp.-2018) \cite{Baynham-2018}. 

\begin{table}[btp]
	\caption{Electric quadrupole moment $\Theta$ (in a.u.), and the scalar ($\alpha_d^{S}$) and tensor ($\alpha_d^{T}$) components of the electric quadrupole polarizabilities (in a.u.) of the $4f^{14}5d~^2D_{3/2}$ state in Yb$^+$. \label{tab:Yb+2D23E2}}
	{\setlength{\tabcolsep}{6pt}
		\begin{tabular}{lcc c}\hline\hline			\addlinespace[0.2cm]
Methods	&	$\Theta$ 	&	$\alpha_q^S$	&	$\alpha_q^T$	\\ \hline\addlinespace[0.2cm]
$2\zeta$ 	&		&		&		\\
e23-CCSD $<$ 10	&	2.001 	&	640 	&	552 	\\
e23-CCSDT  $<$ 10	&	1.985 	&	601 	&	520 	\\ \addlinespace[0.2cm]
$3\zeta$ 	&		&		&		\\
e23-CCSD $<$ 10	&	2.041 	&	740 	&	573 	\\
e33-CCSD $<$ 10	&	2.058 	&		&		\\
e23-CCSD $<$ 10+saug	&	2.037 	&	822 	&	536 	\\
e23-CCSD $<$ 50 	&	2.036 	&	816 	&	641 	\\
$\Delta_{4d}$	&	0.017 	&		&		\\
$\Delta_{saug}$	&	$-0.004$ 	&	82 	&	$-37$ 	\\
$\Delta_{virt}$	&	$-0.005$ 	&	76 	&	68 	\\ \addlinespace[0.2cm]
$4\zeta$ 	&		&		&		\\
e23-CCSD $<$ 10	&	2.017 	&	879 	&	603 	\\
e23-CCSD $<$ 50	&	2.011 	&	805 	&	624 	\\
$\Delta_{basis}$	&	$-0.025$ 	&	$-11$ 	&	$-17$ 	\\ \addlinespace[0.2cm]
Final 	&	1.973(31)	&	672(112)	&	466(80)	\\ \hline\addlinespace[0.2cm]
Exp.-2005 \cite{Schneider-PRL-2005}	&	2.08(11)	&		&		\\
Exp.-2020 \cite{Lange-PRL-2020}	&	1.95(1) 	&		&		\\
MCDHF-2006 \cite{Itano-PRA-2006}	&	2.174	&		&		\\
RCC-2007 \cite{Latha-PRA-2007} 	&	2.157	&		&		\\
RCC-2014 \cite{Nandy-PRA-2014}	&	 2.068(12)	&		&		\\
FSCC-2020 \cite{Guo-CPB-2020}	&	 2.050(30)	&		&		\\\hline\hline
	\end{tabular}}
\end{table}

\begin{table}[btp]
\caption{Static scalar ($\alpha_d^{S}$,) and tensor ($\alpha_d^{T}$,) electric dipole polarizabilities (in a.u.) and the electric quadrupole moment $\Theta$ (in a.u.) of the  $4f^{14}5d~^2D_{5/2}$ state in Yb$^+$. \label{tab:Yb+2D25E2}}
{\setlength{\tabcolsep}{2pt}
		\begin{tabular}{lcc cc}\hline\hline			\addlinespace[0.2cm]
Methods	&	EE	&	 $\alpha_d^S$ 	&	$\alpha_d^T$	&	$\Theta$ 	\\ \hline	\addlinespace[0.2cm]
e23-CCSD $<$ 10 $2\zeta$	&	24848	&	105.13 	&	-86.39 	&	3.170 	\\
e23-CCSD $<$ 10 $3\zeta$	&	24834 	&	94.07 	&	-76.00 	&	3.125 	\\
e23-CCSD $<$ 10 $3\zeta$+saug	&	24809 	&	94.81 	&	-76.74 	&	3.119 	\\
e23-CCSD $<$ 10 $4\zeta$	&	24797 	&	91.27 	&	-74.98 	&	3.092 	\\
$\Delta_{basis}$	&	-37 	&	-2.80 	&	1.02 	&	-0.034 	\\ \addlinespace[0.2cm]
Final	&	24761(73)	& 89(6)	&	$-74(2)$	& 3.06(7)	\\ 
\hline \addlinespace[0.2cm]
NIST \cite{NIST}	&	24332.69	&		&		&		\\
MCDHF-2006 \cite{Itano-PRA-2006}	&		&		&		&	3.244 	\\
RCCSD-2014 \cite{Nandy-PRA-2014}	&		&		&		&	3.116(15)	\\
FSCCSD-2020 \cite{Guo-CPB-2020}	&		&		&		&	3.064(44)	\\ \hline\hline
\end{tabular}}
\end{table}

The $\Theta$ value for the $4f^{14}5d~^2D_{3/2}$ state of Yb$^+$ from our calculations is summarized in Table \ref{tab:Yb+2D23E2}, where it is compared with
other literature values \cite{Schneider-PRL-2005, Lange-PRL-2020, Latha-PRA-2007, Nandy-PRA-2014, Guo-CPB-2020}. Using the 2$\zeta$ basis set, the
calculated quadrupole moment is obtained as $\Theta = 2.001$ a.u. in the e23-CCSD$<$10 approximation. Upon including triple excitations (SDT), the 
value of $\Theta$ decreases slightly to 1.985 a.u., reflecting an improvement due to higher-order electron correlation effects from these excitations. The triple contribution is estimated to be less than 1\%, however, it ensures that the results obtained using e23-CCSDT$<$10 achieve results closer to the experimental value of Ref.~\cite{Lange-PRL-2020}.
With the 3$\zeta$ and 4$\zeta$ basis sets, the results from the e23-CCSD$<$10 approximation show a clear convergence as the basis set size increases. We also incorporated a single-fold diffuse basis set (`saug') in the 3$\zeta$ calculation, which does not have much impact on the final values, but it assures about the robustness in the calculations using the RCC methods. To further validate the calculations, we conducted computations in the e23-SD$<$50 approximation at both the 3$\zeta$ and 4$\zeta$ levels, confirming consistency in the $\Theta$ values with different sizes of basis sets. The final value of the electric quadrupole moment ($\Theta = 1.973(31)$ a.u.) is determined from the e23-SDT$<$10 calculation using the 2$\zeta$ basis set. Adequate corrections are made for residual basis set effects ($\Delta_{basis}$) and higher-order correlation contributions ($\Delta_{4d}$, $\Delta_{saug}$,
and $\Delta_{virt}$). The net uncertainty is again estimated by adding all these corrections in quadrature. This comprehensive treatment of basis set and correlation effects ensures that the final result is more reliable.

Our final $\Theta$ value of the $5D_{3/2}$ state of Yb$^+$ is in excellent agreement with the experimental data; particularly with the value of $\Theta = 1.95(1)$ a.u. reported in 2020 by Lange et al. \cite{Lange-PRL-2020}. The central value from the earlier experiment is 2.08(11) a.u. from Schneider et al. (Exp.-2005)\cite{Schneider-PRL-2005}, which is on slightly higher side. The previous MCDHF calculation (MCDHF-2016) \cite{Itano-PRA-2006}, the RCC calculations (RCC-2007 and RCC-2014) \cite{Latha-PRA-2007,Nandy-PRA-2014} , and the FSCC calculation (FSCCSD-2020) \cite{Guo-CPB-2020} slightly overestimated this value. Similarly, the $\alpha_q^S$ and $\alpha_q^T$ values for the $4f^{14}5d~^2D_{3/2}$ state of Yb$^+$ are also summarized in Table \ref{tab:Yb+2D23E2}, using the same strategy discussed for the $\Theta$ value above. To our knowledge, there are no other literature values available to compare with our results.

In Table \ref{tab:Yb+2D25E2}, we present results from the e32-CCSD calculation for EE, $\alpha_d^S$, $\alpha_d^T$, and $\Theta$ for the $5D_{5/2}$ state of Yb$^+$. As can be seen,  the energy decreases slightly, ranging from 24848 to 24797, when the basis set size increases from $2\zeta$ to $4\zeta$, and slightly larger than the NIST value \cite{NIST}. Correspondingly, the values for $\alpha_d^S$, $\alpha_d^T$, and $\Theta$ decrease with increasing basis set size, showing a clear tendency towards convergence. We estimate the correction $\Delta_{basis}$ due to the basis set in terms of the differences between the results for $X=3\xi$ and $X=4\zeta$, and the final results are taken as the values are taken as sum of values from $4\xi$ and $\Delta_{basis}$. To account for uncertainties due to triple excitation contributions, we estimate the uncertainty by applying a factor of 2 to $\Delta_{basis}$ from the analyses of the results for the $5D_{3/2}$ state and to be on safer side. We compare the $\Theta$ value with previous results from MCDHF \cite{Itano-PRA-2006} and RCCSD \cite{Nandy-PRA-2014}, which show slightly lower value but it is in agreement with the Fock-space based RCC calculation of Ref. \cite{Guo-CPB-2020}. 

\section*{Summary}

This study employs the general-order relativistic coupled-cluster method at the SD (single and double excitations) and SDT (single, double, and triple excitations) levels to calculate critical electric-field response properties—$\alpha_d$ (dipole polarizability), $\Theta$ (quadrupole moment), and $\alpha_q$ (tensor polarizability)—for the $6s~^2S_{1/2} \rightarrow 5d~^2D_{3/2,5/2}$ clock transitions in Yb$^+$. Our results for the $6s~^2S_{1/2} \rightarrow 5d~^2D_{3/2}$ clock transition underscore the vital roles of triple excitation contributions in improving the accuracy of these calculations, particularly for $\alpha_d$ and $\Theta$, compared to the analogous clock transition in the lighter Ca$^+$ ion.

This work also highlights and addresses discrepancies in previous theoretical and experimental results for $\alpha_d$, $\Theta$, and $\alpha_q$. By incorporating triple excitations, we provide refined calculations that bridge these gaps, enhance theoretical benchmarks, and improve computational accuracy. These advancements guide experimental setups by helping control electric fields and temperature distributions to minimize field-induced energy shifts. Moreover, new experimental measurements complement these theoretical efforts, establishing benchmarks for high-precision studies and advancing clock performance.

Based on our findings in Yb$^+$, we recommend the scalar ground state dipole polarizability as $\alpha_d^S = 63.53(35)$ and the scalar and tensor dipole polarizability values for the excited $4f^{14}5d~^2D_{3/2}$ state as $\alpha_d^S = 101(4)$ and $\alpha_d^T =-72(5)$ (all in atomic units). These values yield a differential scalar dipole polarizability of $\Delta \alpha_d^S = 38(4)$~a.u. for the $6s~^2S_{1/2} \rightarrow 5d~^2D_{3/2}$ clock transition of Yb$^+$, in excellent agreement with the experimental result from Ref.~\cite{Baynham-2018}. Additionally, the quadrupole moment $\Theta = 1.973(31)$~a.u. for the $4f^{14}5d~^2D_{3/2}$ state aligns well with the recent experimental value reported in Ref.~\cite{Lange-PRL-2021}.

\section*{ACKNOWLEDGMENTS}
	
This work is supported by The National Key Research and Development Program of China (2021YFA1402104), Innovation Program for Quantum Science and Technology (2021ZD0300901), and a Project supported by the Space Application System of China Manned Space Program. The work of BKS is supported by ANRF grant no. CRG/2023/002558 and Department of Space, Government of India for financial supports, and he acknowledges use of the
ParamVikram-1000 HPC cluster of Physical Research Laboratory (PRL), Ahmedabad, Gujarat, India.


\begin{thebibliography}{}
	
\bibitem{Taylor-PRA-1997}
P. Taylor, M. Roberts, S. V. Gateva-Kostova, R. B. M. Clarke, G. P. Barwood, W. R. C. Rowley, and P. Gill, Phys. Rev. A {\bf 56}, 2699 (1997).

\bibitem{Roberts-PRA-1999}
M. Roberts, P. Taylor, S. V. Gateva-Kostova, R. B. M. Clarke, W. R. C. Rowley, P. Gill, Phys. Rev. A {\bf 60}, 2867 (1999).	

\bibitem{Webster-IEEE-2010}
S. A. Webster, R. M. Godun, S. A. King, G. Huang, B. Walton, V. Tsatourian, H. S. Margolis, S. Lea and P. Gill, IEEE Trans. on Ultrasonics, Ferroelectrics, and Frequency Control {\bf57}, 3 (2010).

\bibitem{Huntemann-PRA-2014}
C. Tamm, N. Huntemann, B. Lipphardt, V. Gerginov, N. Nemitz, M. Kazda, S. Weyers and E. Peik, Phys. Rev. A {\bf 89}, 023820 (2014).

\bibitem{Godun-PRL-2014}
R. M. Godun, P. B. R. Nisbet-Jones, J. M. Jones, S. A. King, L. A. M. Johnson, H. S. Margolis, K. Szymaniec, S. N. Lea, K. Bongs, and P. Gill, Phys. Rev. Lett. {\bf 113}, 210801 (2014).

\bibitem{Huntemann-PRL-2014}
N. Huntemann, B. Lipphardt, Chr. Tamm, V. Gerginov, S. Weyers, and E. Peik, Phys. Rev. Lett. {\bf 113}, 210802 (2014).

\bibitem{Filzinger-PRL-2023}
M. Filzinger, S. D\"{o}rscher, R. Lange, J. Klose , M. Steinel , E. Benkler , E. Peik , C. Lisdat , and N. Huntemann, Phys. Rev. Lett. {\bf 130}, 253001 (2023).

\bibitem{Lange-PRL-2021}
R. Lange, N. Huntemann, J. M. Rahm, C. Sanner, H. Shao, B. Lipphardt, C. Tamm, S. Weyers, and E. Peik, Phys. Rev. Lett. {\bf 126}, 011102 (2021).

\bibitem{Sanner-Nature-2019}	
C. Sanner, N. Huntemann, R. Lange, C. Tamm, E. Peik, M. S. Safronova, and S. G. Porsev, Nature (London) {\bf 567}, 204 (2019).

\bibitem{Tofful-Metro-2024}
A. Tofful, C. F. A. Baynham, E. A. Curtis, A. O. Parsons, B. I. Robertson, M. Schioppo, J. Tunesi, H. S. Margolis, R. J. Hendricks, J. Whale, R. C. Thompson, and R. M. Godun, Metrologia  {\bf 61}, 045001 (2024).

\bibitem{Flambaum-CJP-2009}
V. V. Flambaum and V. A. Dzuba, Can. J. Phys. {\bf 87}, 25 (2009).

\bibitem{Schneider-PRL-2005}	
T. Schneider, E. Peik, and Chr. Tamm, Phys. Rev. Lett., {\bf 94}, 230801 (2005).	

\bibitem{Baynham-2018}
C. F. A. Baynham, E. A. Curtis, R. M. Godun, J. M. Jones, P. B. R. Nisbet-Jones, P. E. G. Baird, K. Bongs, P. Gill, arXiv:1801.10134v3, (2020).

\bibitem{Lange-PRL-2020}
R. Lange, N. Huntemann, C. Sanner, H. Shao, B. Lipphardt , Chr. Tamm, and E. Peik, Phys. Rev. Lett. {\bf 125}, 143201 (2020).

\bibitem{Migdalek-1982}
J. Migdalek, J. Quant. Spectrosc. Radiat. Transf. {\bf 28}, 61(1982).
 		
\bibitem{Lea-2006}
S. N. Lea, S. A. Webster, and G. P. Barwood, Proceedings of the 20th EFTF, 302, (2006).

\bibitem{Safronova-PRA-2009}
 U. I. Safronova and M. S. Safronova, Phys. Rev. A {\bf 79}, 022512 (2009).

\bibitem{Porsev-PRA-2012}
S. G. Porsev, M. S. Safronova, and M. G. Kozlov, Phys. Rev. A {\bf 86}, 022504 (2012).

\bibitem{Roy-2017}
A. Roy, S. De, Bindiya Arora and B. K. Sahoo, J. Phys. B: At. Mol. Opt. Phys. {\bf 50}, 205201 (2017).

\bibitem{Itano-PRA-2006}	
Wayne M. Itano, Phys. Rev. A {\bf 73}, 022510 (2006).

\bibitem{Latha-PRA-2007}
K. V. P. Latha, C. Sur, R. K. Chaudhuri, B. P. Das, and D. Mukherjee, Phys. Rev. A {\bf 76}, 062508 (2007).

\bibitem{Nandy-PRA-2014}
D. K. Nandy and B. K. Sahoo, Phys. Rev. A {\bf 90}, 050503(R) (2014).

\bibitem{Guo-CPB-2020}
X. T. Guo, Y. M. Yu, Y. Liu, and B. B. Suo, Chin. Phys. B {\bf 29} 053101 (2020).		

\bibitem{Chen-CPB-2023}
T. Chen, L. Wu, R.-K. Zhang, Y.-B. Tang, J. Jiang, and C.-Z. Dong, Chin. Phys. B {\bf 32}, 053206 (2023).

\bibitem{Sahoo-PRA-2011}
B. K. Sahoo and B. P. Das, Phys. Rev. A {\bf 84}, 010502(R) (2011).

\bibitem{Blythe-JPB-2003}
P. J. Blythe, S. A. Webster, K. Hosaka and P. Gill, J. Phys. B {\bf 36}, 981 (2003).

\bibitem{Bishop-TCA-1991}
R. F. Bishop, Theoretica Chimica Acta {\bf 80}, 95 (1991).

\bibitem{Dirac}
DIRAC, a relativistic ab initio electronic structure program, Release DIRAC22 (2022), written by H. J. {\relax Aa. Jensen}, R. Bast, A. S. P. Gomes, T. Saue and L. Visscher, with contributions from I. A. Aucar, V. Bakken, C. Chibueze, J. Creutzberg, K. G. Dyall,
S. Dubillard, U. Ekstr{\"o}m, E. Eliav, T. Enevoldsen, E. Fa{\ss}hauer, T. Fleig, O. Fossgaard,
L. Halbert, E. D. Hedeg{\aa}rd, T. Helgaker, B. Helmich--Paris, J. Henriksson, M. van Horn,
M. Ilia{\v{s}}, Ch. R. Jacob, S. Knecht, S. Komorovsk{\'y}, O. Kullie, J. K. L{\ae}rdahl, C. V. Larsen,
Y. S. Lee, N. H. List, H. S. Nataraj, M. K. Nayak, P. Norman, G. Olejniczak,
J. Olsen, J. M. H. Olsen, A. Papadopoulos, Y. C. Park, J. K. Pedersen, M. Pernpointner,
J. V. Pototschnig, R. Di Remigio, M. Repisky, K. Ruud, P. Sa{\l}ek, B. Schimmelpfennig,
B. Senjean, A. Shee, J. Sikkema, A. Sunaga, A. J. Thorvaldsen, J. Thyssen,
J. van Stralen, M. L. Vidal, S. Villaume, O. Visser, T. Winther, S. Yamamoto and X. Yuan
(available at \url{http://dx.doi.org/10.5281/zenodo.6010450},
see also  \url{http://www.diracprogram.org}).

\bibitem{Dirac2020}
T. Saue, R. Bast, A. S. P. Gomes, H. J. A. Jensen, L. Visscher, I. A. Aucar, R. Di Remigio, K. G. Dyall, E. Eliav, E. Fa{\ss}hauer, T. Fleig, L. Halbert, E. D. Hedeg{\aa}rd, B. Helmich-Paris, M. Ilia\v{s}, C. R. Jacob, S. Knecht, J. K. Laerdahl, M. L. Vidal, M. K. Nayak, M. Olejniczak, J. M. H. Olsen, M. Pernpointner, B. Senjean, A. Shee, A. Sunaga, and J. N. P. van Stralen, J. Chem. Phys. {\bf 152}, 204104 (2020).



\bibitem{Kallay-JCP-2002}
M. K\'{a}llay,, P. G. Szalay, and P. R. Surj\'{a}n, J. Chem. Phys. {\bf 117}, 980 (2002).

\bibitem{Kallay-JCP-2004}
M. K\'{a}llay, and J. Gauss, J. Chem. Phys. {\bf 121}, 9257 (2004).

\bibitem{MRCC}
Mrcc, a quantum chemical program suite written by M. K\'{a}llay, Z. Rolik, J. Csontos, I. Ladj\'{a}nszki, L. Szegedy, B. Lad\'{o}czki, and G. Samu. See also Z. Rolik, L. Szegedy, I. Ladj\'{a}nszki, B. Lad\'{o}czki, and M. K\'{a}llay, J. Chem. Phys. {\bf 139}, 094105 (2013), as well as: www.mrcc.hu.

\bibitem{Dyall-Ca}
K.G. Dyall and A.S.P. Gomes, unpublished.

\bibitem{Bindiya-PRA-2007}
B. Arora, M. S. Safronova, and C. W. Clark, Phys. Rev. A {\bf 76}, 064501 (2007).

\bibitem{Sahoo-PRA-2009}
B. K. Sahoo, B. P. Das, and D. Mukherjee, Phys. Rev. A, {\bf79}, 052511 (2009).

\bibitem{Huang-PRA-2019}
Y. Huang, H. Guan, M. Zeng, L. Tang, and K. Gao, Phys. Rev. A, {\bf99}, 011401(R) (2019).

\bibitem{Theodosiou-PRA-1995}
C. E. Theodosiou, L. J. Curtis, and C. A. Nicolaides, Phys.Rev. A {\bf52}, 3677 (1995).

\bibitem{Barklem-PRA-1998}
P. S. Barklem and B. J. O’Mara, Mon. Not. R. Astron. Soc. {\bf300}, 863 (1998).

\bibitem{Gomes-TCA-2010}
Andr\'{e} S. P. Gomes, K. G. Dyall, L. Visscher, Theor. Chem. Acc. {\bf 127}, 369 (2010).

\bibitem{NIST}
A. Kramida, Y. Ralchenko, J. Reader, and NIST ASD Team https://physics.nist.gov/asd.
	
\end{thebibliography}
\end{document}